\@twosidetrue
\@twocolumntrue
\@mparswitchtrue
\def\ds@draft{\overfullrule 5pt}
\def\ds@twocolumn{\@twocolumntrue}
\def\ds@onecolumn{\@twocolumnfalse}
\newif\ifSFB@landscape
\def\ds@landscape{\SFB@landscapetrue}
\newif\ifSFB@galley
\def\ds@galley{\SFB@galleytrue}
\newif\ifSFB@referee
\def\ds@referee{%
 \SFB@refereetrue
 \@twocolumnfalse
}
\@options
\ifSFB@referee
 
\else
 
\fi
\if@twocolumn
  \def\@normalsize{\@setsize\normalsize{11pt}\ixpt\@ixpt
   \abovedisplayskip 6pt plus 2pt minus 2pt
   \belowdisplayskip \abovedisplayskip
   \abovedisplayshortskip 6pt plus 2pt
   \belowdisplayshortskip \abovedisplayshortskip
   \let\@listi\@listI}
 \else
 \ifSFB@referee
  \def\@normalsize{\@setsize\normalsize{14pt}\xiipt\@xiipt
   \abovedisplayskip 4pt plus 1pt minus 1pt
   \belowdisplayskip \abovedisplayskip
   \abovedisplayshortskip 4pt plus 1pt
   \belowdisplayshortskip \abovedisplayshortskip
   \let\@listi\@listI}
 \else
  \def\@normalsize{\@setsize\normalsize{12pt}\ixpt\@ixpt
   \abovedisplayskip 4pt plus 1pt minus 1pt
   \belowdisplayskip \abovedisplayskip
   \abovedisplayshortskip 4pt plus 1pt
   \belowdisplayshortskip \abovedisplayshortskip
   \let\@listi\@listI}
 \fi
\fi
\def\small{\@setsize\small{10pt}\viiipt\@viiipt
 \abovedisplayskip 4pt plus 1pt minus 1pt
 \belowdisplayskip \abovedisplayskip
 \abovedisplayshortskip 4pt plus 1pt
 \belowdisplayshortskip \abovedisplayshortskip
 \def\@listi{\leftmargin\leftmargini
  \topsep 2pt plus 1pt minus 1pt
  \parsep \z@
  \itemsep 2pt}}
\def\footnotesize{\@setsize\footnotesize{10pt}\viiipt\@viiipt
 \abovedisplayskip 4pt plus 1pt minus 1pt
 \belowdisplayskip \abovedisplayskip
 \abovedisplayshortskip 4pt plus 1pt
 \belowdisplayshortskip \abovedisplayshortskip
 \def\@listi{\leftmargin\leftmargini
  \topsep 2pt plus 1pt minus 1pt
  \parsep \z@
  \itemsep 2pt}}
\def\scriptsize{\@setsize\scriptsize{8pt}\viipt\@viipt}
\def\tiny{\@setsize\tiny{6pt}\vpt\@vpt}
\if@twocolumn
  \def\large{\@setsize\large{11pt}\xpt\@xpt}
 \else
  \def\large{\@setsize\large{12pt}\xpt\@xpt}
 \fi
\def\Large{\@setsize\Large{14pt}\xiipt\@xiipt}
\def\LARGE{\@setsize\LARGE{17pt}\xivpt\@xivpt}
\def\huge{\@setsize\huge{20pt}\xviipt\@xviipt}
\def\Huge{\@setsize\huge{25pt}\xxpt\@xxpt}
\normalsize

\if@twocolumn
\else
 \ifSFB@referee
  \oddsidemargin \z@  \evensidemargin \z@
 \else
 \fi
\fi
 \topmargin \z@
\newdimen\SFB@measure
\textwidth \SFB@measure
\ifSFB@landscape
 \textwidth \textheight
 \textheight \SFB@measure
\fi
\ifSFB@referee
\fi
\skip\footins 19.5pt plus 12pt minus 1pt

\parskip \z@ plus .1pt
\@beginparpenalty -\@lowpenalty
\@endparpenalty -\@lowpenalty
\@itempenalty -\@lowpenalty
\newcounter{part}
\newcounter {section}
\newcounter {subsection}[section]
\newcounter {subsubsection}[subsection]
\newcounter {paragraph}[subsubsection]
\newcounter {subparagraph}[paragraph]
\def\thepart          {\arabic{part}}
\def\thesection       {\arabic{section}}

\def\part{\par \addvspace{4ex}\@afterindentfalse
 \secdef\@part\@spart}
\def\@part[#1]#2{\ifnum \c@secnumdepth >\m@ne
  \refstepcounter{part}
  \addcontentsline{toc}{part}{Part \thepart: #1}
 \else \addcontentsline{toc}{part}{#1}
 \fi
 {\parindent 0pt \raggedright
  \ifnum \c@secnumdepth >\m@ne
   \large\rm PART
   \ifcase\thepart \or ONE \or TWO \or THREE \or FOUR \or FIVE
    \or SIX \or SEVEN \or EIGHT \or NINE \or TEN \else \fi
   \par \nobreak
  \fi
  \LARGE \rm #2 \markboth{}{}\par }
 \nobreak \vskip 3ex \@afterheading}
\def\@spart#1{{\parindent 0pt \raggedright
  \LARGE \rm #1\par}
 \nobreak \vskip 3ex \@afterheading}

\def\section{\@startsection {section}{1}{\z@}
 {-24pt plus -12pt minus -1pt}
 {6pt}
 {\SFB@hangraggedright\normalsize\bf}}
\def\subsection{\@startsection{subsection}{2}{\z@}
 {-18pt plus -9pt minus -1pt}
 {6pt}
 {\SFB@hangraggedright\large\bf}}
\def\subsubsection{\@startsection{subsubsection}{3}{\z@}
 {-18pt plus -9pt minus -1pt}
 {6pt}
 {\SFB@hangraggedright\normalsize\it}}
\def\paragraph{\@startsection{paragraph}{4}{\z@}
 {12pt plus 2.25pt minus 1pt}{-0.5em}{\normalsize\bf}}
\def\subparagraph{\@startsection{subparagraph}{5}{\parindent}
 {12pt plus 2.25pt minus 1pt}{-0.5em}{\normalsize\it}}
\def\SFB@hangraggedright{\rightskip\@flushglue \let\\=\newline}
\def\@sect#1#2#3#4#5#6[#7]#8{%
 \ifnum #2>\c@secnumdepth
  \def\@svsec{}%
 \else
  \refstepcounter{#1}
  \ifnum #2=\@ne
   \ifSFB@appendix \edef\@svsec{}%
             \else \edef\@svsec{\csname the#1\endcsname\hskip 1em}%
   \fi
  \else \edef\@svsec{\csname the#1\endcsname\hskip 1em}%
  \fi
 \fi
 \@tempskipa #5\relax
 \ifdim \@tempskipa>\z@
  \begingroup #6\relax
   \ifnum #2=\@ne
    \ifSFB@appendix
     \@hangfrom{\hskip #3\relax\@svsec}{\interlinepenalty \@M
      APPENDIX \csname the#1\endcsname:\hskip 0.5em\uppercase{#8}\par}%
    \else
     \@hangfrom{\hskip #3\relax\@svsec}{\interlinepenalty \@M
      \uppercase{#8}\par}%
    \fi
   \else
    \@hangfrom{\hskip #3\relax\@svsec}{\interlinepenalty \@M #8\par}%
   \fi
  \endgroup
  \csname #1mark\endcsname{#7}%
  \addcontentsline{toc}{#1}{\ifnum #2>\c@secnumdepth \else
   \protect\numberline{\csname the#1\endcsname}\fi #7}%
 \else
  \def\@svsechd{#6\hskip #3\@svsec \ifnum #2=\@ne\uppercase{#8}\else #8\fi
  \csname #1mark\endcsname{#7}
  \addcontentsline{toc}{#1}{\ifnum #2>\c@secnumdepth \else
   \protect\numberline{\csname the#1\endcsname}\fi#7}}%
 \fi
 \@xsect{#5}}

\newif\ifSFB@appendix
\def\appendix{\par
 \SFB@appendixtrue
 \setcounter{section}{0}
 \def\thesection{A\arabic{section}}
 \setcounter{equation}{0}
 \def\theequation{A\arabic{equation}}
 \setcounter{figure}{0}
 \def\thefigure{A\@arabic\c@figure}
 \setcounter{table}{0}
 \def\thetable{A\@arabic\c@table}
}

\newskip\@indentskip
\newskip\smallindent
\newskip\@footindent
\newskip\@leftskip
\@footindent=\smallindent
\@leftskip=\z@

\leftmargini   \@indentskip
\leftmargin\leftmargini
\labelwidth\leftmargini\advance\labelwidth-\labelsep
\def\makeRRlabel#1{\hss\llap{#1}}
\def\@listI{\leftmargin\leftmargini
 \parsep \z@
 \topsep 6pt plus 1pt minus 1pt
 \itemsep \z@ plus .1pt
}
\let\@listi\@listI
\@listi
\def\@listii{\leftmargin\leftmarginii
 \labelwidth\leftmarginii\advance\labelwidth-\labelsep
 \topsep 6pt plus 1pt minus 1pt
 \parsep \z@
 \itemsep \z@ plus .1pt
}
\def\@listiii{\leftmargin\leftmarginiii
 \labelwidth\leftmarginiii\advance\labelwidth-\labelsep
 \topsep 6pt plus 1pt minus 1pt
 \parsep \z@
 \partopsep \z@
 \itemsep \topsep
}
\def\@listiv{\leftmargin\leftmarginiv
 \labelwidth\leftmarginiv\advance\labelwidth-\labelsep
}
\def\@listv{\leftmargin\leftmarginv
 \labelwidth\leftmarginv\advance\labelwidth-\labelsep
}
\def\@listvi{\leftmargin\leftmarginvi
 \labelwidth\leftmarginvi\advance\labelwidth-\labelsep
}
\def\itemize{\ifnum \@itemdepth >3 \@toodeep
  \else \advance\@itemdepth \@ne
   \edef\@itemitem{labelitem\romannumeral\the\@itemdepth}%
   \list{\csname\@itemitem\endcsname}%
    {\let\makelabel\makeRRlabel}%
  \fi}

\def\enumerate{\ifnum \@enumdepth >3 \@toodeep \else
  \advance\@enumdepth \@ne
  \edef\@enumctr{enum\romannumeral\the\@enumdepth}%
 \fi
 \@ifnextchar [{\@enumeratetwo}{\@enumerateone}%
}
\def\@enumeratetwo[#1]{%
 \list{\csname label\@enumctr\endcsname}%
  {\settowidth\labelwidth{[#1]}
   \leftmargin\labelwidth \advance\leftmargin\labelsep
   \usecounter{\@enumctr}
   \let\makelabel\makeRRlabel}
}
\def\@enumerateone{%
 \list{\csname label\@enumctr\endcsname}%
  {\usecounter{\@enumctr}
   \let\makelabel\makeRRlabel}}

\def\theenumi{(\roman{enumi})}

\def\theenumii{(\alph{enumii})}
\def\p@enumii{\theenumi}

\def\theenumiii{(\arabic{enumiii})}
\def\p@enumiii{\theenumi(\theenumii)}

\def\p@enumiv{\p@enumiii\theenumiii}
\def\description{\list{}{\labelwidth\z@ \itemindent-\leftmargin
  \leftmargin 1em
  \itemindent-1em
}}

\def\verse{\let\\=\@centercr
 \list{}{\itemsep\z@
  \itemindent -\@indentskip
  \listparindent \itemindent
  \rightmargin\leftmargin
  \advance\leftmargin \@indentskip}\item[]}

\def\quotation{\list{}{\listparindent \smallindent
  \leftmargin\z@\rightmargin\leftmargin
  \parsep 0pt plus 1pt}\item[]\small}

\def\quote{\list{}{\leftmargin\z@\rightmargin\leftmargin}\item[]\small}

\def\@begintheorem#1#2{\rm \trivlist \item[\hskip \labelsep{\bf #1\ #2.}]}
\def\@opargbegintheorem#1#2#3{\rm \trivlist
  \item[\hskip \labelsep{\bf #1\ #2.\ (#3)}]}
\def\@endtheorem{\endtrivlist}
\@namedef{proof*}{\rm \trivlist \item[\hskip \labelsep{\it Proof.}]}
\@namedef{endproof*}{\endtrivlist}

\def\titlepage{\@restonecolfalse\if@twocolumn\@restonecoltrue\onecolumn
  \else \newpage \fi \thispagestyle{empty}\c@page\z@}
\def\endtitlepage{\if@restonecol\twocolumn \else \newpage \fi}

\def\tabular{\def\@halignto{}
 \def\hline{\noalign{\ifnum0=`}\fi
  \vskip 3pt
  \hrule \@height \arrayrulewidth
  \vskip 3pt
  \futurelet \@tempa\@xhline}
 \def\fullhline{\noalign{\ifnum0=`}\fi
  \vskip 3pt
  \hrule \@height \arrayrulewidth
  \vskip 3pt
  \futurelet \@tempa\@xhline}
 \def\@xhline{\ifx\@tempa\hline
   \vskip -6pt
   \vskip \doublerulesep
  \fi
  \ifnum0=`{\fi}}
  \def\@arrayrule{\@addtopreamble{\hskip -.5\arrayrulewidth
                                  \hskip .5\arrayrulewidth}}
\@tabular
}
\tabbingsep \labelsep

\skip\@mpfootins = \skip\footins

\fboxrule = \arrayrulewidth

\def\maketitle{\par
 \begingroup
  \if@twocolumn
   \twocolumn[\vspace*{17pt}\@maketitle]
  \else
   \newpage
   \global\@topnum\z@
   \@maketitle
  \fi
  \thispagestyle{titlepage}
 \endgroup
 \let\maketitle\relax
 \let\@maketitle\relax
 \gdef\@author{}
 \gdef\@title{}
 \let\thanks\relax
}
\def\and{\end{author@tabular}\vskip 6pt\par
 \begin{author@tabular}[t]{@{}l@{}}}
\def\@maketitle{\newpage
 \vspace*{7pt}
 {\raggedright \sloppy
  {\huge \bf \@title \par}
  \vskip 23pt
  {\LARGE
   \begin{author@tabular}[t]{@{}l@{}}\@author
   \end{author@tabular}\par}
  \vskip 22pt
 }
 \par\noindent
 {\small \@date \par}
 \vskip 22pt
}
\def\abstract{\if@twocolumn
  \start@SFBbox\@abstract
 \else
  \@abstract
 \fi}
\def\endabstract{\if@twocolumn
   \endlist\finish@SFBbox
 \else
  \endlist
 \fi}
\def\@abstract{\list{}{\leftmargin 10.5pc\rightmargin\z@
  \parsep 0pt plus 1pt}\item[]\normalsize{\bf ABSTRACT}\\\large} 
\newif\ifSFB@keywords
\def\keywords{\if@twocolumn
  \start@SFBbox\@keywords
 \else
  \@keywords
 \fi
}
\def\@keywords{\list{}{\leftmargin 10.5pc\rightmargin\z@
  \parsep 0pt plus 1pt}\item[]\large{\bf Key words: }}
\def\endkeywords{\if@twocolumn
  \endlist\addvspace{37pt}\finish@SFBbox
 \else
  \endlist
 \fi
 \@thanks
 \gdef\@thanks{}
 \SFB@keywordstrue
}
\def\nokeywords{\ifSFB@keywords\else
 \if@twocolumn \start@SFBbox\addvspace{37pt}\finish@SFBbox \fi
 \@thanks
 \gdef\@thanks{}\fi
}

\def\author@tabular{\def\@halignto{}\@authortable}
\let\endauthor@tabular=\endtabular
\def\author@tabcrone{{\ifnum0=`}\fi\@xtabularcr[-7pt]\small\it
 \let\\=\author@tabcrtwo\ignorespaces}
\def\author@tabcrtwo{{\ifnum0=`}\fi\@xtabularcr[-7pt]\small\it
 \let\\=\author@tabcrtwo\ignorespaces}
\def\@authortable{\leavevmode \hbox \bgroup $\let\@acol\@tabacol
 \let\@classz\@tabclassz \let\@classiv\@tabclassiv
 \let\\=\author@tabcrone \ignorespaces \@tabarray}

\def\start@SFBbox{\@next\@currbox\@freelist{}{}%
 \global\setbox\@currbox
 \vbox\bgroup
  \hsize \textwidth
  \@parboxrestore
}
\def\finish@SFBbox{\par\vskip -\dbltextfloatsep
  \egroup
  \global\count\@currbox\tw@
  \global\@dbltopnum\@ne
  \global\@dbltoproom\maxdimen\@addtodblcol
  \global\vsize\@colht
  \global\@colroom\@colht
}

\mark{{}{}}
\gdef\@author{\mbox{}}
\def\author{\@ifnextchar [{\@authortwo}{\@authorone}}
\def\@authortwo[#1]#2{\gdef\@author{#2}\gdef\@shortauthor{#1}}
\def\@authorone#1{\gdef\@author{#1}\gdef\@shortauthor{#1}}
\gdef\@shortauthor{}
\gdef\@title{\mbox{}}
\def\title{\@ifnextchar [{\@titletwo}{\@titleone}}
\def\@titletwo[#1]#2{\gdef\@title{#2}\gdef\@shorttitle{#1}}
\def\@titleone#1{\gdef\@title{#1}\gdef\@shorttitle{#1}}
\gdef\@shorttitle{}
\def\volume#1{\gdef\@volume{#1}}
\gdef\@volume{000}
\def\microfiche#1{\gdef\@microfiche{#1}}
\gdef\@microfiche{}
\def\pagerange#1{\gdef\@pagerange{#1}}
\gdef\@pagerange{000--000}
\def\journal#1{\gdef\@journal{#1}}
\gdef\@journal{{Mon.\ Not.\ R.\ Astron.\ Soc.} {\bf \@volume}, \@pagerange\
  (\number\year) \@microfiche}
\def\ps@headings{\let\@mkboth\markboth
 \def\@oddhead{\Large \hfill \it \@shorttitle \hspace{1.5em}\rm \thepage}
 \def\@oddfoot{}
 \def\@evenhead{\Large \thepage \hspace{1.5em}\it \@shortauthor \hfill}
 \def\@evenfoot{}
 \def\sectionmark##1{\markboth{##1}{}}
 \def\subsectionmark##1{\markright{##1}}}
\def\ps@myheadings{\let\@mkboth\@gobbletwo
 \def\@oddhead{\Large \it \rightmark \hfill \rm \thepage}
 \def\@oddfoot{}
 \def\@evenhead{\Large \it \leftmark \hfill \rm \thepage}
 \def\@evenfoot{}
 \def\sectionmark##1{}
 \def\subsectionmark##1{}}
\def\ps@titlepage{\let\@mkboth\@gobbletwo
 \def\@oddhead{\footnotesize\@journal\hfill}
 \def\@oddfoot{}
 \def\@evenhead{\footnotesize\@journal\hfill}
 \def\@evenfoot{}
 \def\sectionmark##1{}
 \def\subsectionmark##1{}}

\def\@pnumwidth{1.55em}
\def\@tocrmarg {2.55em}
\def\@dotsep{4.5}
\def\@undottedtocline#1#2#3#4#5{\ifnum #1>\c@tocdepth
 \else
  \vskip \z@ plus .2pt
  {\hangindent #2\relax
   \rightskip \@tocrmarg \parfillskip -\rightskip
   \parindent #2\relax \@afterindenttrue
   \interlinepenalty\@M \leavevmode
   \@tempdima #3\relax #4\nobreak \hfill \nobreak
   \hbox to\@pnumwidth{\hfil\rm \ }\par}\fi}
\def\tableofcontents{\@restonecolfalse
 \if@twocolumn\@restonecoltrue\onecolumn\fi
 \section*{CONTENTS} \@starttoc{toc}
 \if@restonecol\twocolumn\fi \par\vspace{12pt}}
\def\l@part#1#2{\addpenalty{-\@highpenalty}
 \addvspace{2.25em plus 1pt}
 \begingroup
  \parindent \z@ \rightskip \@pnumwidth
  \parfillskip -\@pnumwidth
  {\normalsize\rm
   \leavevmode \hspace*{3pc}
   #1\hfil \hbox to\@pnumwidth{\hss \ }}\par
   \nobreak \global\@nobreaktrue
   \everypar{\global\@nobreakfalse\everypar{}}\endgroup}
\def\l@section#1#2{\addpenalty{\@secpenalty}
 \@tempdima 1.5em
 \begingroup
  \parindent \z@ \rightskip \@pnumwidth
  \parfillskip -\@pnumwidth \rm \leavevmode
  \advance\leftskip\@tempdima \hskip -\leftskip
  #1\nobreak\hfil \nobreak\hbox to\@pnumwidth{\hss \ }\par
 \endgroup}
\def\l@subsection{\@undottedtocline{2}{1.5em}{2.3em}}
\def\l@subsubsection{\@undottedtocline{3}{3.8em}{3.2em}}
\def\l@paragraph{\@undottedtocline{4}{7.0em}{4.1em}}
\def\l@subparagraph{\@undottedtocline{5}{10em}{5em}}
\def\listoffigures{\@restonecolfalse
 \if@twocolumn\@restonecoltrue\onecolumn\fi
 \section*{LIST OF FIGURES\@mkboth{LIST OF FIGURES}{LIST OF FIGURES}}
 \@starttoc{lof} \if@restonecol\twocolumn\fi}
\def\l@figure{\@undottedtocline{1}{1.5em}{2.3em}}
\def\listoftables{\@restonecolfalse
 \if@twocolumn\@restonecoltrue\onecolumn\fi
 \section*{LIST OF TABLES\@mkboth{LIST OF TABLES}{LIST OF TABLES}}
 \@starttoc{lot} \if@restonecol\twocolumn\fi}
\let\l@table\l@figure

\def\thebibliography#1{\section*{REFERENCES}
 \addcontentsline{toc}{section}{REFERENCES}
 \list{}{\labelwidth\z@
         \leftmargin 1.5em
	 \itemsep \z@
	 \itemindent-\leftmargin}
 \small\raggedright
 \parindent\z@
 \parskip\z@ plus .1pt\relax
 \def\newblock{\hskip .11em plus .33em minus .07em}
 \sloppy\clubpenalty4000\widowpenalty4000
 \sfcode`\.=1000\relax
}

\def\@biblabel#1{\hspace*{\labelsep}[#1]}

\newif\if@restonecol
\def\theindex{\section*{INDEX}
 \addcontentsline{toc}{section}{INDEX}
 \footnotesize \parindent\z@ \parskip\z@ plus .1pt\relax
 \let\item\@idxitem}
\def\@idxitem{\par\hangindent 1em}

\def\endtheindex{\if@restonecol\onecolumn\else\clearpage\fi}

\def\footnoterule{\kern-3\p@ \hrule width 12pc height \z@ \kern 3\p@}

\def\@fnsymbol#1{\ifcase#1\or \mbox{$\star$}\or \dagger\or \ddagger\or
   \S \or \P \or \|\or **\or \dagger\dagger
   \or \ddagger\ddagger\or \S\S\or \P\P\or \|\|\else ***
   \fi\relax}

\long\def\@makefntext#1{\parindent 1em\noindent
  $^{\@thefnmark}$\hspace{4pt}#1}

\newcounter{table}
\def\thetable{\@arabic\c@table}
\def\fps@table{tbp}
\def\ftype@table{1}
\def\ext@table{lot}
\def\fnum@table{Table \thetable}
\def\table{\let\@makecaption=\SFB@maketablecaption\@float{table}}
\let\endtable\end@float
\@namedef{table*}{\let\@makecaption=\SFB@maketablecaption\@dblfloat{table}}
\@namedef{endtable*}{\end@dblfloat}

\newcounter{figure}
\def\thefigure{\@arabic\c@figure}
\def\fps@figure{tbp}
\def\ftype@figure{2}
\def\ext@figure{lof}
\def\fnum@figure{Figure \thefigure}
\def\figure{\let\@makecaption=\SFB@makefigurecaption\@float{figure}}
\let\endfigure\end@float
\@namedef{figure*}{\let\@makecaption=\SFB@makefigurecaption\@dblfloat{figure}}
\@namedef{endfigure*}{\end@dblfloat}

\long\def\SFB@makefigurecaption#1#2{\vskip 6pt
 \setbox\@tempboxa\hbox{\small{\bf #1.} #2}
 \ifdim \wd\@tempboxa >\hsize
  \small{\bf #1.} #2\par
 \else
  \hbox to\hsize{\hfil\box\@tempboxa\hfil}
 \fi
 \vskip 6pt
}
\long\def\SFB@maketablecaption#1#2{\vskip 6pt
 \setbox\@tempboxa\hbox{\small{\bf #1.} #2}
 \ifdim \wd\@tempboxa >\hsize
  \small{\bf #1.} #2\par
 \else
  \hbox to\hsize{\box\@tempboxa\hfill}
 \fi
 \vskip 6pt
}

\def\caption{\@ifstar{\SFB@caption\@captype}%
 {\refstepcounter\@captype \@dblarg{\@caption\@captype}}%
}
\long\def\SFB@caption#1#2{
 \begingroup
  \@parboxrestore
  \normalsize
  \@makecaption{\csname fnum@#1\endcsname}{\ignorespaces #2}\par
 \endgroup}

\def\@cite#1#2{(#1\if@tempswa , #2\fi)}
\def\@biblabel#1{}

\newlength{\bibhang}
\def\@citex[#1]#2{\if@filesw\immediate\write\@auxout{\string\citation{#2}}\fi
  \def\@citea{}\@cite{\@for\@citeb:=#2\do
    {\@citea\def\@citea{; }\@ifundefined
       {b@\@citeb}{{\bf ?}\@warning
       {Citation `\@citeb' on page \thepage \space undefined}}%
{\csname b@\@citeb\endcsname}}}{#1}}
\let\@internalcite\cite
\def\cite{\def\citename##1{##1}\@internalcite}
\def\shortcite{\def\citename##1{}\@internalcite}

\def\[{\relax\ifmmode\@badmath\else\begin{trivlist}\item[]\leavevmode
  \hbox to\linewidth\bgroup$
  \displaystyle
  \hskip\mathindent\bgroup\fi}

\def\]{\relax\ifmmode \egroup $\hfil
       \egroup \end{trivlist}\else \@badmath \fi}

\def\equation{\refstepcounter{equation}\trivlist \item[]\leavevmode
  \hbox to\linewidth\bgroup $
  \displaystyle
\hskip\mathindent}

\def\endequation{$\hfil
           \displaywidth\linewidth\@eqnnum\egroup \endtrivlist}

\def\eqnarray{\stepcounter{equation}\let\@currentlabel=\theequation
\global\@eqnswtrue
\global\@eqcnt\z@\tabskip\mathindent\let\\=\@eqncr
\abovedisplayskip\topsep\ifvmode\advance\abovedisplayskip\partopsep\fi
\belowdisplayskip\abovedisplayskip
\belowdisplayshortskip\abovedisplayskip
\abovedisplayshortskip\abovedisplayskip
$$\halign
to \linewidth\bgroup\@eqnsel\hskip\@centering$\displaystyle\tabskip\z@
  {##}$&\global\@eqcnt\@ne \hskip 2\arraycolsep \hfil${##}$\hfil
  &\global\@eqcnt\tw@ \hskip 2\arraycolsep $\displaystyle{##}$\hfil
   \tabskip\@centering&\llap{##}\tabskip\z@\cr}

\def\endeqnarray{\@@eqncr\egroup
 \global\advance\c@equation\m@ne$$\global\@ignoretrue}

\newdimen\mathindent
\mathindent = \z@

\def\today{\number\day\ \ifcase\month\or
  January\or February\or March\or April\or May\or June\or
  July\or August\or September\or October\or November\or December
 \fi \ \number\year}

\ps@headings
\ifSFB@galley
 \ps@empty
\fi
\ifSFB@referee
\fi
\if@twocolumn
 \twocolumn
\else
 \onecolumn
\fi
\documentstyle{mn}

\title{Gravitational Waves and the Polarisation of the
Cosmic Microwave Background}
\author[R. A. Frewin, A. G. Polnarev and P. Coles]
       {R.A. Frewin, A.G. Polnarev and P. Coles\\
        Astronomy Unit, School of Mathematical Sciences,
Queen Mary and Westfield College, Mile End Road, London E1 4NS}
\date{Accepted 1993 ???? ???;
      Received 1993 ???? ???;
      in original form 1993 ???? ??}

\begin{document}

\maketitle

 \begin{abstract}
 We discuss the influence of gravitational waves (GWs)
upon the polarisation
of the Cosmic Microwave Background Radiation (CMBR). We show how to
compute the {\em rms} temperature anisotropy and polarisation of the CMBR
induced by GWs of arbitrary wavelength. We find that the ratio of
polarisation, $\Pi$, to
anisotropy,  $A$, can be as large as $\sim 40$\%, but is
sensitively dependent upon the GW spectrum and the cosmological
ionisation history. We argue that CMBR polarisation measurements
can provide useful constraints on cosmological models.

 \end{abstract}

\section{Introduction}

The recent discovery by the COBE team of angular fluctuations in the
sky temperature of the Cosmic Microwave Background Radiation (CMBR)
is of profound importance for theories of the origin of galaxies and
large-scale structures in the Universe \cite{cobe}.
If the observed temperature anisotropy
is interpreted as being due to fluctuations
in the density of the Universe at early times then, together with
measurements of present-day galaxy clustering, it imposes strong
constraints on the primordial fluctuation spectrum and the nature of
and dark matter \cite{ebw,trr}.
These constraints will be further strengthened by measurements
of temperature anisotropy on angular scales smaller than those
probed by COBE
\cite{belm}.

Compared to the enormous observational effort that has been directed
at the search for anisotropy in the temperature of the CMBR on the sky,
relatively little attention has been paid to analogous fluctuations in its
{\em polarisation}. Atmospheric contributions to the sky
temperature at microwave frequencies are not polarised so it is possible
to make measurements of polarisation from the ground. To this extent
at least, the observational task is less problematic than trying to
detect temperature anisotropy \cite{p88},
though there are of course other problems of experimental design
\cite{cf78,ls79,n79,lms,p88}.
On the other hand,
in ``standard'' models of galaxy formation via gravitational instability
from primordial adiabatic density inhomogeneities,
the polarisation is expected to be much smaller than the
temperature anisotropy \cite{k83,be84,nn93}. There are situations, however,
when the ratio of polarisation to
temperature fluctuations can be non-negligible. Rees \shortcite{r68}
showed that an axisymmetric anisotropic cosmological
expansion should induce a significant large-scale
polarisation of the CMBR. This work was subsequently
extended (and corrected) by Basko \& Polnarev \shortcite{bp80},
who obtained an exact solution
to for the polarisation anisotropy produced in a flat triaxial anisotropic
cosmological model; see also \cite{ns80,tm84,t85}.
Polarisation fluctuations are also induced if there exists a background
of tensor perturbations in the metric (i.e. gravitational waves, hereafter
GW's) at the time of recombination. Indeed, one can regard the triaxial
cosmological model mentioned above \cite{bp80}
as being the superposition of an infinitely--long wavelength
GW on an homogeneous and isotropic background space--time; the axisymmetric
case corresponds to a scalar perturbation to a homogeneous and
isotropic model. Polnarev (1985) subsequently showed how
to calculate the polarisation anisotropy due to scattering
of radiation by electrons in  the presence of a single
GW and obtained some analytic formulae for various
limiting cases of gravitational wavelength and recombination history.

The possible existence of a cosmological (stochastic)
GW background has been discussed for some time
\cite{b75,g75,st79,c80}. More recently
it has been realised that inflationary models of the early
Universe produce stochastic GWs with characteristic spectra
\cite{rsv,aw84,st85,lm85,ah86,a88,sahn}.
Although these tensor perturbation
modes have no influence on the formation of cosmic structures, they do
generate anisotropies in the  CMBR temperature \cite{sw,d69,gz78,am78}.
This has led a number of authors recently to produce
inflationary models in which a significant fraction of the
temperature anisotropy detected by COBE would be due to
(tensor) GWs rather than (scalar) density perturbations
\cite{dhsst,ll92,lc92,lmm,s92,cbdes}. If this is the
case then the constraints imposed by COBE upon models of structure formation
are considerably altered. Moreover, there is a possibility that one might
be able to use the ratio of amplitudes of tensor and scalar modes to
reconstruct the shape of the effective potential of the scalar
field responsible for driving inflation \cite{cklla,ckllb}.
Measurements of
the CMBR temperature anisotropy at a single angular scale do not
allow one to discriminate between contributions from scalar and tensor
modes and, although the angular dependence of the temperature anisotropy
is different in the scalar and tensor cases, there are problems in
using information from smaller angular scales than COBE
because of the possibility
that reionisation might mask the behaviour of the primordial fluctuations
\cite{belm}.

In this paper we shall argue that CMBR
polarisation measurements can supply
important information about the existence of a significant cosmological
GW wave background and also about the ionisation history of the Universe.
We shall concentrate on explaining the basic physics behind the
generation of polarisation; computations of detailed statistical
properties will be deferred to a later paper.
Throughout this paper we shall assume that the background cosmology
is described by a flat metric:
\begin{equation}
ds^{2}= -dt^{2} + a(t)^{2}d{\bf x}^{2} =
a(\eta)^{2} \left( -d\eta^{2} + d{\bf x}^{2} \right),
\end{equation}
where $t$ is cosmological proper time, $a$ is the scale factor and
$\eta$ is conformal time ($d\eta=dt/a(t)$); the ${\bf x}$ are comoving
coordinates; the speed of light is unity.

\section{Radiative Transfer with GWs}

The essence of our problem is to calculate the effect of gravitational
radiation upon the transfer of electromagnetic radiation through
the period of hydrogen recombination and photon decoupling. To
proceed we therefore need to consider the effect of
Thomson-scattering upon the polarisation of electromagnetic
radiation. An unpolarised beam of
radiation picks up linear polarisation during Thomson scattering; the maximum
polarisation is 100 \% for radiation scattered perpendicular to the
incident beam. If the incoming radiation is isotropic in the
rest frame of the scattering particle, the scattered radiation is
unpolarised. The same is true if the radiation has a dipole anisotropy.
If there is a quadrupole anisotropy, however, there should be a net
polarisation of the scattered radiation. Such a quadrupole can
be caused by a density perturbation or, as we shall investigate in this
paper, a gravitational wave.
We adopt the formalism suggested by \cite{ch} and construct
a vector ${\bf n}$ with components $n_{r}$, $n_{l}$ and $n_{u}$ related
to the usual Stokes parameters. Here $n_r+n_l=n$, the total photon
occupation number. (The polarisation tensor of the radiation, $\pi_{ij}$,
has off-diagonal elements equal to $n_{u}/2$ and diagonal elements
equal to $n_{l}$ and $n_{r}$ respectively.)
In the presence of a single gravitational wave, the
components of ${\bf n}$ are functions of: (i) conformal time $\eta$;
(ii) comoving spatial coordinates ${\bf x}$; (iii) photon frequency $\nu$;
(iv) the polar angle, $\theta=\cos^{-1} \mu$, between ${\bf \hat{q}}$ (a unit
vector
in the direction of photon propagation) and ${\bf \hat{k}}$ (a
unit vector in along the GW); (v) the azimuthal angle, $\phi$, between the
projection of ${\bf \hat{q}}$
onto a plane perpendicular to ${\bf k}$ and a unit vector
perpendicular to ${\bf k}$ derived from the GW polarisation tensor
\cite{pol}.

The equation of radiative transfer can be written in terms of the
vector ${\bf n}$, as follows:
\begin{equation}
\frac{\partial {\bf n}}{\partial \eta} + {\bf \hat{q}} \cdot
\frac{\partial {\bf n}}{\partial {\bf x}} =
- \frac{\partial {\bf n}}{\partial\nu} \frac{\partial \nu}{\partial \eta}
-\sigma_{T}N_e a \left[ {\bf n} - {\bf I}({\bf n}) \right],
\end{equation}
where
\begin{equation}
{\bf I}({\bf n}) = \frac{1}{4\pi}
\int_{-1}^{+1} \int_{0}^{2\pi} {\bf P}(\mu,\phi,\mu'\phi')\,
{\bf n} \,d\mu' d\phi';
\end{equation}
$\sigma_{T}$ is the usual Thomson scattering cross-section,
$N_e$ is the comoving number-density of free electrons and
${\bf P}$ is the scattering matrix
which is described by \cite{ch},
and is given explicitly in terms of these variables by \cite{pol}.
The important term in this context is the effect of the gravitational
wave in shifting the photon frequencies via the first term on the
right hand side of equation (2).

If the Universe is flat and filled with pressureless matter
the appropriate linearised Einstein equations admit a solution for
tensor metric perturbations $h_{\alpha}^{\beta}$ which, for
a single wave, can be written in the form:
\begin{equation}
h_{\alpha}^{\beta} = h \epsilon_{\alpha}^{\beta} \exp [-i{\bf k}\cdot{\bf x}
+i\omega(k)\eta].
\end{equation}
The wavenumber $k$ is defined such that the physical wavelength
$\lambda=2\pi a/k$ and, because $c=1$, we have $\omega(k)=k$;
$\epsilon_{\alpha}^{\beta}$ is the GW polarisation tensor. The
geodesic equation in the perturbed metric
yields
\begin{equation}
\frac{d\nu}{d\eta} = \frac{\nu}{2}(1-\mu^{2}) e^{-i{\bf k}\cdot
{\bf x}} \cos 2\phi  \frac{d}{d\eta} \left(h e^{ik\eta}\right).
\end{equation}
In the unperturbed case ($h=0$), the solution
to (2) is simply ${\bf n} = n_0(1,1,0)$.
To obtain the solution to first order in $h$ it
proves convenient \cite{bp80,pol} to transform to an alternative
set of symbolic vectors which reduces (2) to a system of
integro-differential equations which has an exact analytical solution
in the limit $k\rightarrow 0$. The procedure for doing this will not
be described here; see \cite{pol}. In terms of new variables
$\alpha(\eta,\nu,\mu)$, $\beta(\eta,\nu,\mu)$ and $\xi=\alpha+\beta$
we obtain
\begin{equation}
\dot{\beta} + [q- ik\mu]\beta  = F
\end{equation}
\begin{equation}
\dot{\xi}+[q-ik\mu]\xi = H,
\end{equation}
where $q=\sigma_T N_e a$ and
\begin{equation}
F(\eta) = \frac{3q}{16} \int_{-1}^{+1}
\left[ \left(1+\mu'^{2}\right)^{2}\beta - \frac{1}{2}\xi\left(1-\mu'^{2}
\right)^{2} \right]d\mu',
\end{equation}
\begin{equation}
H(\eta) = \frac{C(k)}{k^{3}}\frac{\partial}{\partial\eta}
\left(\frac{1}{\eta}\frac{\partial}{\partial\eta} \frac{\sin k\eta}{\eta}
\right).
\end{equation}
Note that we use a different definition of $h$ compared to
Polnarev (1985);
$C(k)$ is related to the GW spectrum \cite{st85}.
We shall concentrate in this paper on the evaluation of the
{\em rms} polarisation and anisotropy induced by GWs of a given
wavenumber in stochastic superposition:
\begin{equation}
\Pi_{k}=\langle \Pi_k^{2} \rangle^{1/2} = \left[
\int_{-1}^{+1} |\beta|^{2} (1+\mu^{2})^{2}d\mu \right]^{1/2};
\end{equation}
\begin{equation}
A_{k}=\langle A_k^{2} \rangle^{1/2} = \left[
\int_{-1}^{+1} |\xi|^{2} (1-\mu^{2})^{2}d\mu \right]^{1/2}.
\end{equation}

Before displaying our results, it is useful to re-cap the
analytical results \cite{pol}. One can find a solution
to the stationary case with $H=\bar{H}={\rm const.}$
and $q=\bar{q}={\rm const}$. Here $\Pi_k$ and $A_k$ are of the same order,
at least for small
angular scales $\theta \leq (2q/k)^{1/2}$.
This solution corresponds to the case
prior to recombination. To go further one can assume an instantaneous
recombination such that the stationary solution applies until
$\eta=\eta_r$ and thereafter $q=0$. Since there is no more
scattering after $\eta_r$ the polarisation remains unchanged between
$\eta_r$ and the present epoch. The anisotropy, however, grows because
of the Sachs-Wolfe effect which does not involve scattering. The
ratio of polarisation to anisotropy is therefore expected
to be small in such a situation.
However, in realistic cosmological models, we
do not expect recombination to be instantaneous. If there is
an extended period of ionisation then scattering can, in principle,
generate an interestingly large value of $\Pi_k$.

\section{Results}

The solution of the equations (6)--(9) can be expressed
formally as
\begin{equation}
\left( \begin{array}{c} \beta \\ \xi \end{array}\right)
= e^{\tau+ik\mu\eta} \int_{0}^{\eta}d\eta'
\left( \begin{array}{c} F(\eta') \\ H(\eta') \end{array}\right)
e^{-\tau(\eta')-ik\mu\eta'},
\end{equation}
where the optical depth,
$\tau=\int_{\eta}^{1} q(\eta') d\eta'$, so that $\tau=0$ when
$\eta=1$. To find solutions for the {\em present}
mean square anisotropy and polarisation, one simply evaluates
(10) \& (11) at $\eta=1$. The results can thus be expressed as integrals
over $\eta$:
\begin{equation}
\langle \Pi_k^{2} \rangle = \frac{32}{3}
\int_{0}^{1} d\eta \frac{\bar{F}^{2}(\eta) e^{-2\tau(\eta)}}{q(\eta)} +
\int_{0}^{1}d\eta  \bar{F}(\eta)F_0(\eta)
\end{equation}
\begin{equation}
\langle A_k^{2} \rangle = 2\int_{0}^{1}d\eta H(\eta)  e^{-\tau(\eta)}
F_0(\eta),
\end{equation}
where $\bar{F}(\eta)=F(\eta)e^{-\tau(\eta)}$.
The function $F_0(\eta)$ is given by
\begin{equation}
F_0 = \int_{0}^{\eta} d\eta' H(\eta') e^{-\tau(\eta')} K_{-} (\eta,\eta')
\end{equation}
and $\bar{F}(\eta)$ must be obtained by solving the integral equation
\begin{equation}
\bar{F} = \frac{3q}{16} e^{\tau(\eta)} \left\{ \int_{0}^{\eta}
\bar{F}(\eta') K_{+} (\eta,\eta')d\eta' - \frac{1}{2}F_{0}(\eta')
d\eta' \right\}.
\end{equation}
The function $K_{\pm}$ used in these expressions is just
\begin{equation}
K_{\pm}(\eta,\eta') = \int_{-1}^{+1} d\mu \left(1\pm \mu^{2}\right)^{2}
e^{ik\mu(\eta-\eta')}.
\end{equation}
Performing the integrals over $\mu$ first allows us to extend
the work of \cite{pol} to arbitrary $q$ and $k$ whilst keeping
the number of numerical integrations required to a minimum.
A convenient parametrisation is $q(\eta)=q_r \chi(\eta)\eta^{-4}$
where $\chi$ is the fractional ionisation; $q_0 \simeq 0.14\Omega_b h$,
where $h$ is the Hubble parameter in units of 100 km s$^{-1}$ Mpc$^{-1}$.
For small $\eta$, $\chi=1$; we adopt the following flexible
illustrative model for the variation of $\chi$ through
recombination:
\begin{equation}
\chi(\eta) = \left\{
\begin{array}{cc}
1-\frac{2(\eta-\eta_r)^{2}}{\Delta^{2}} & \eta< \eta_r + \frac{\Delta}{2}\\
\chi_0 + \frac{2(\eta-\eta_r-\Delta)^{2}}{\Delta^{2}} &
\eta_r + \frac{\Delta}{2} < \eta < \eta_r + \Delta\\
\chi_0 & \eta > \eta_r + \Delta. \end{array} \right.
\end{equation}
Here $\Delta$ parametrises the duration of recombination and $\chi_0$
is the residual ionisation. The standard picture of recombination has
$\eta_r=\eta_s \simeq 0.026$, $\Delta=\Delta_s\simeq 0.05$
and $\chi_0 \simeq 3 \times 10^{-5}
(\Omega_0/\Omega_b h)$; standard cosmological nucleosynthesis requires
$0.010 < \Omega_b h < 0.032$ (Olive et al. 1990); we take $\Omega_0$
to be unity throughout these calculations. The advantage of the simple
model (18) is that it is easy to integrate and allows us simply to
assess the effect upon the level of polarisation of changes in the
parameters $\chi_0$, $\Delta$ and $\eta_r$. To reduce the parameter
space somewhat, we fix the optical depth at the end of recombination
$\eta=\eta_r+\Delta$ to be equal for all the models we consider; the
standard value is $\tau=0.07$. In this way have only two independent
parameters which we take to be $\Delta$ and $\eta_r$. As an
extreme example we also consider the case of no recombination at all,
$\chi(\eta)=1$.

We shall look at $\Pi_k$ and $A_k$ as functions of $k$ in the following
series of figures. In Figure 1 we study the anisotropy produced as a
function of $k$ for different
models of the ionisation history. The GWs all have an arbitrary
initial amplitude independent of $k$ for this and the subsequent figures.
The trend with recombination model is straightforward:
the more extended the period of ionisation, the smaller
the anisotropy produced at large $k$. This is due to the blurring
effect of the finite width of the last scattering surface. In
the extreme case of no-recombination, there is severe suppression
of the small-scale anisotropy.

\begin{figure}
\centering
\vspace{5cm}
\caption{Anisotropy as a function of $k$
for different models of the ionisation history. Models
are shown with:  $\eta_r=\eta_s$, $\Delta=\Delta_s$ (standard);
$\eta_r=2\eta_s$, $\Delta=\Delta_s$; $\eta_r=\eta_s$, $\Delta=5\Delta_s$;
$\eta_r=\eta_s$, $\Delta=100\Delta_s$. A model with no recombination is shown
for comparison. }
\label{fig1}
\end{figure}

Figure 2 shows the polarisation for the same set of models as Fig. 1.
The most important point here is that the maximum polarisation shows the
opposite trend to the anisotropy: the longer the period of ionisation, the
higher is the peak polarisation. The peak wavelength  also increases
as the width of the last scattering surfaces increases: scattering can
occur later, when the horizon size is larger. The oscillations in
polarisation
for large $k$ are caused by resonances between the GW wavelength and the
width of the last scattering surface.

\begin{figure}
\centering
\vspace{5cm}
\caption{As Figure 1, but showing the polarisation as a function of $k$
for the different models.}
\label{fig2}
\end{figure}

Figures 1 and 2 show very clearly the basic physics in operation
during the production of a polarised CMBR. Because of the
arbitrary scaling, however, they do not represent quantities
that can be compared to observation. For this, we need to look
at the ratio of polarisation to anisotropy. This poses some
problems. The ratio $\Pi_k/A_k$
would correspond to the ratio of polarisation to anisotropy
observed for a $\delta$--function spectrum. For broader
spectra -- particularly the very flat spectra typically predicted
in inflation \cite{rsv,aw84,st85,lm85,ah86,a88,sahn}
-- a more relevant characterisation would be
the ratio of total polarisation
$\Pi =(\int \langle \Pi_k^{2} \rangle dk/k)^{1/2}$ to {\em total} anisotropy
$A=(\int \langle A_k^{2} \rangle dk/k)^{1/2}$, which would
generally be smaller than $\Pi_k/A_k$ evaluated at a single point.
Furthermore, any given experiment will observe some particular angular
scale on the sky which would correspond to a weighted sum of
contributions from all $k$. To explore systematically the space
of beam-widths and GW spectra is beyond the scope of this
paper; we shall restrict ourselves to showing $\Pi_k/A_k$ for a few
examples to show when this ratio can be large (see Figure 3).
The ratio $\Pi_k/A_k$ has a maximum value of around 10\%
for the standard
model, increasing to over 40\% for the no recombination case.
Note, however, that the ratio of {\em total} polarisation
to {\em total} anisotropy
(integrated over a flat spectrum) is indeed very much smaller
than this: $\Pi/A\simeq 0.3$\% for standard recombination and
$\Pi/A\simeq
3.7$\% for no recombination. Clearly the superposition of GWs with
different wavelengths leads to a large reduction in the observable
polarisation compared to the $\delta$ function case.

\begin{figure}
\centering
\vspace{5cm}
\caption{As Figure 3, but showing the ratio of polarisation to anisotropy
as a function of $k$ for the different models.}
\label{fig3}
\end{figure}

\section{Discussion and Conclusions}

We have seen that, in certain conditions, a stochastic GW background
can lead to a significant polarisation of the CMBR. The level
of polarisation is strongly dependent upon the GW spectrum: it is
high for a $\delta$--function, but much lower for a flat spectrum.
Whether the level is high enough to be observed would depend on
the GW spectrum, the ionisation history and the experimental beamwidth.
We shall explore this parameter space more systematically in a forthcoming
paper; preliminary analysis suggests that experiments capable of
detecting  $\Pi/A\leq 10$\% would be needed to provide useful data.

In the inflationary models there will be both scalar and tensor
contributions to both polarisation and anisotropy. The values we have
obtained for the tensor perturbations
are larger than those usually
quoted for scalar modes \cite{k83,be84,nn93} on large scales, but scalar
perturbations have means other
than the Sachs--Wolfe effect for inducing anisotropy and polarisation
(e.g. streaming motions and the Silk effect). These mechanisms depend
sensitively upon the dark matter and normalisation of the density
fluctuations, which makes a full calculation of the contribution to
the polarisation from both modes difficult. Suppose that
the total anisotropy $A$ and polarisation
$\Pi$ includes both scalar and tensor contributions:
$A=A_T+A_S$ and $\Pi=\Pi_{T}+\Pi_S$. Now if tensor modes contribute
a fraction $f$ of the total anisotropy then the overall
ratio of polarisation to anisotropy is just
\begin{equation}
\left(\frac{\Pi}{A}\right) = f \left( \frac{\Pi}{A} \right)_{T} +
(1-f) \left(\frac{\Pi}{A} \right)_{S}
\end{equation}
(assuming tensor and scalar modes add independently, as they should
in linear theory). Only if $f$ is significant
and the ionisation history is such that $(\Pi/A)_{T}$ is large
can one expect there to be a significant alteration in the overall
ratio of polarisation to anisotropy compared to the standard case.
In the models discussed by Crittenden
et al. (1993b) the polarisation induced by the tensor modes
is usually smaller than the scalar contribution: at best it is
comparable.  If one does not know {\em a priori} how much of the anisotropy
is produced by tensor modes, it would be very difficult to
use polarisation measurements to disentangle the contribution
in such models. However, these authors considered only a small subset
of inflationary models. Models can be produced which yield a
much larger value of $f$ than they considered:
it is possible to have $f\simeq 0.5$
without violating constraints on the fluctuation power spectrum
\cite{dhsst,ll92,lc92,lmm,s92}. We shall give specific predictions
for particular inflationary spectra in a forthcoming paper.

It is worth mentioning, however, that even if the scalar and tensor
contributions to the total polarisation are comparable in terms
of their {\em rms} values, one might still be able to discriminate
between them. For a start, the autocorrelation functions
of the temperature pattern
(not calculated in this paper) will be different in the two cases,
because the correlation angle is determined by physical length scales
which are different for scalar and tensor modes, as can be seen from
Figs 1 \& 2. Recently, Naselsky \& Novikov (1993) have argued
that oscillatory features (similar to `Sakharov' oscillations)
in the power spectrum of fluctuations
produced by adiabatic scalar fluctuations could be a powerful
cosmological probe. The detailed spatial distribution
of polarisation and anisotropy could also be a sensitive
discriminant. For example, the relative positions of `hotspots' of
temperature and polarisation are different in the tensor and
scalar case. The simplest way to characterise this would be
to calculate the cross-correlation between polarisation and
anisotropy maps. We shall return to these ideas in future work.

\subsection*{Acknowledgments}
RAF receives an SERC postgraduate studentship. This work was
partially supported by SERC under the QMW Rolling Theory Grant
GR/ H09454. We thank Andy Liddle and Paul Steinhardt for valuable
comments. While this work was nearing completion, we received
preprints by \cite{cds} and \cite{nn93},
in which similar questions are considered.

\end{document}